\pdfoutput=1
\documentclass[twoside,english,prl, nobibnotes, nofootinbib, twocolumn, showkeys, showpacs, balancelastpage, floatfix]{revtex4}
\usepackage[T1]{fontenc}
\usepackage[latin9]{inputenc}
\usepackage[a4paper]{geometry}
\usepackage{babel}
\usepackage{amsmath}
\usepackage{amssymb}
\usepackage{graphicx}
\usepackage{esint}
\usepackage[unicode=true,
 bookmarks=true,bookmarksnumbered=false,bookmarksopen=false,
 breaklinks=false,pdfborder={0 0 1},backref=false,colorlinks=false]
 {hyperref}
\hypersetup{pdftitle={A Theory of Quantum Observation and the Emergence of the Born Rule},
 pdfauthor={Andreas Oliver Tell}}

\makeatletter
\@ifundefined{textcolor}{}
{%
 \definecolor{BLACK}{gray}{0}
 \definecolor{WHITE}{gray}{1}
 \definecolor{RED}{rgb}{1,0,0}
 \definecolor{GREEN}{rgb}{0,1,0}
 \definecolor{BLUE}{rgb}{0,0,1}
 \definecolor{CYAN}{cmyk}{1,0,0,0}
 \definecolor{MAGENTA}{cmyk}{0,1,0,0}
 \definecolor{YELLOW}{cmyk}{0,0,1,0}
}

\usepackage{braket}

\makeatother

\begin{document}

\title{A Theory of Quantum Observation and the Emergence of the Born Rule}

\author{Andreas O. Tell}

\email{aotell@ifqt.de}

\date{\inputencoding{latin1}\today}
\selectlanguage{english}%
\begin{abstract}
A realist description of our universe requires a twofold concept of locality. 
On one hand, there are the strictly Einstein-local interactions which generate
the time evolution. On the other hand, the quantum state space calls for
a non-local description of multi-particle states. This article uses a behavioristic 
approach to argue, that an observer in a universe like this has to rely on local 
interactions to learn about its properties and behavior. 
Such an observer is fundamentally restricted in his ability to understand and 
structurally reconstruct the individual local physical universe. We argue, that 
this reconstruction based on dynamically available information is the defining 
process of observation in quantum theory. The observer-centric view of the global 
quantum dynamics is shown to be non-unitary and non-linear in general, even if 
the universe itself evolves unitarily.

Interactions with massless free particles are found to have great influence on observation,
because of their special role in the light-cone structure of an Einstein-local universe. 
For a specific scattering process with a photon of unknown state, the observed
 outcome  can be subjectively random and follow the Born statistic, while the output state 
is really determined by the photon polarization. Based on this result, a theory of quantum 
measurement is formulated, which describes a measurement device as a cascade of scattering events.
\end{abstract}



\maketitle

\section{Introduction}

Quantum theory is an enormously successful theory. We can use it to
predict the behavior of nature very precisely up to the limits of measurement.
This is even more astounding as we have almost no understanding of what ``measuring'' 
really means. Quantum theory equips us with an algorithm \citep{Wallace2007} to describe, 
how the possible outcomes of a measurement can be calculated and predicted statistically. 
However, the measurement postulate of traditional\footnote{Traditional quantum theory 
specifically means Neumann-Dirac here.} quantum  theory does not define, what a measurement 
instrument physically is, nor how  it could  possibly function. Even worse, it appears, 
that the process of measurement is  not compatible with the smooth unitary evolution so 
fundamental to quantum theory.

A lot of effort has been placed into resolving this \emph{measurement problem}
and has resulted in a collection of interpretations of quantum theory
\citep{Schlosshauer2003,Wallace2007}. Despite significant progress
in some areas like decoherence theory, a generally accepted solution
to the measurement problem is not currently known. From a realist perspective,
one of the most important aspects of such a solution has to be a derivation of
the statistical law of observations, the Born rule \citep{Born1926},
from fundamental properties of bare quantum theory.
While such derivations have been attempted\footnote{There are many approaches
that cannot be listed here. Please find an overview of the most relevant ones in 
 \citep{Wallace2007}.} \citep{Everett1957,SchlosshauerFine2005,Zurek2004, Zurek2007, Zurek2009},
they usually involve non-obvious assumptions\footnote{Like the postulation of
some ad-hoc probability measure or something equivalent.} or have been shown to rely on
circular or in other ways problematic arguments \citep{Adler2003,Baker2006,Barnum1999,%
Dawid2013,Finkelstein2009,Hemmo2007,Jarlskog2011,Kent1990,Kent2009,Mohrhoff2004,Myrvold2005,Rae2008}.

This article attempts to derive the discontinuous and random, yet stable
results of observation as stated in the measurement postulate only
from generally accepted first principles. This is done by assuming
the position of an observer, who studies his environment in order to collect information
about its behavior. Because the behavior is uniquely and completely determined by the well
defined dynamics of the system, the result qualifies for a true realistic theory of quantum
observation. That includes the possibility of experimental verification.

Everything we know about the universe is a result of our interaction
with it, while being a part of it. As obvious as this statement may
seem, it is of fundamental importance to our understanding of the
way we describe the universe, create physical theories and interpret
them. Even in the non-local setting of quantum theory, all interactions
are subject to locality. In a relativistic universe, an observer
is limited to his light cone and does not have access, in the form
of interaction, to the complete state of the universe.

The idea of considering limited knowledge is not fundamentally new. The information
available to the observer and the change of knowledge during measurement are also subject
of the Copenhagen Interpretation, more precisely the quantum subjectivism
proposed by Werner Heisenberg \citep{Heisenberg1979}. However, his idea has not evolved
into a quantitative theory and it does not state the exact nature of the lack of knowledge.

Obtaining a quantitative characterization of the information available to the
local observer corresponds to describing the state of the universe, that
the observer would be able to reconstruct, based on its behavior, just from
interaction without any additional assumptions.

To demonstrate the key idea, consider a simplified relativistic universe,
containing just two (well localized) particles and an observer located close to one
of these particles. The particles are supposed to have interacted a long time ago
and are spatially well separated, to make sure that they cannot interact for another
sufficiently long time. The observer, only being able to interact with one of the
particles, is not in the position to tell if the two particles are
entangled. In fact, he is unable to tell if a second particle exists
at all. Based on the system's behavior and not making unfounded assumptions,
the observer's description of the universe must not contain the other particle.
The state he reconstructs, or assumes, if you will, is specifically not an improper
 mixed state, because that too would require knowledge about the existence of the other
particle. He is left with the option of describing a single particle with a pure state,
as we will discuss later in more detail.

A different, but similar, scenario is created by the interaction of
the local system with free massless particles. Instead of a distant
particle, imagine a photon, that has just been emitted into the environment.
The information it carries is inaccessible as soon, as it leaves the
observer's horizon. Incoming photons also only become accessible the
very instant, they reach the observer. The moment of interaction
creates a discontinuity in the observer's best guess for the state
of the universe. The transport of information to and from the observer
by interaction with the ambient radiation field also creates a source of randomness.
The unknown state of the incoming radiation influences the local system
in a way unpredictable to the local observer. The effect of the random
perturbation is not entirely unpredictable however. It is restricted
by the nature of the interaction, and shaped by the way the system
couples to the radiation field. This opens the door for a statistical
distribution, that correlates with the previously known state of the
system, just like the Born distribution does. We will see, that
the relevant no-go arguments are not applicable, because the linear structure
of the state space is compromised by the local information constraint.

\section{The Postulates}

In order to build the theory of local observation onto firm grounds,
we use the following postulates. They formalize the assumptions
used in the derivations presented and are well founded in what we already
know about quantum theory. Most importantly, they do not introduce any new
kind of structure.
\begin{enumerate}
\item The evolution of the state of the universe is unitary and generated
only by interactions within the universe.
\item The interactions in the universe are local and in agreement with Special
Relativity.
\item The observer mechanism is part of the universe, and its relevant parts are physically
realized within a finite region of spacetime.
\item The description of the universe, as constructed by the observer, is based
on his local state history, without any additional assumptions.
\end{enumerate}

As observers, we do not have access to the state of the
universe and a reduction of information from top to bottom might
seem unnatural. We have to rely on Occam's principle to argue, that
a theory, which predicts local observations from an unknown global mechanisms,
is valid, if the assumptions made on the global scale are very plain
and simple. This is true for the postulates listed above and therefore
supports the postulate of a global unitary evolution.

Postulate 4 not only restricts the description of the universe to
the information processing happening in the observer system. It also
defines, what we can regard to be physically distinguishable properties
of the universe. Properties, that no imaginable observer can distinguish,
cannot be part of the ontological description of the universe.
It is only the dynamical sequence of the states of the universe---its behavior---, that
results in an emergent physical reality.

In addition to these postulates, we still have to make one assumption
about the universe to be able to derive the Born rule. That is the existence
of a free massless neutral spin-1 boson field, that interacts with the observer.
The photon takes this role in our universe.

The rest of this article uses these postulates to derive fundamental
properties of all possible descriptions of the universe by an observer.
There is no possible strategy, a local mechanism could follow to circumvent
these properties. They are \emph{binding} and must be taken into
account when discussing the apprearance of the universe from an observer
point of view.

\section{Local State Behavior}

\subsection{The state space of the universe}
\label{sec:StatUni}

Our first postulate demands unitary evolution of the state of the
universe, but it does not specify the state space. Instead, we will
try to construct the state space from first principles.

The state space must be able to hold unitary representations of the
observed symmetries of the universe. It must also be able to encode
superposition and interference of states. These requirements are fulfilled
by a complex vector space with a sesquilinear inner product,
where the inner product allows unitary transforms to be defined.
It is not entirely unreasonable to assume, that the state space is
finite dimensional%
\footnote{The representation theory of the Lorentz group $\mathrm{SO}(3,1)$
shows, that a unitary representation requires an infinite dimensional
space. So a finite dimensional quantum state space implies, that the Lorentz
group can only be realized approximately.%
}, but it is mathematically more convenient to allow for an infinite number
of dimensions. In this case, the convergence of the inner product must be listed
as an additional requirement. And we arrive at the well known separable
Hilbert space $\mathcal{H}$ of square integrable functions.

Another possible state space, which meets all requirements, is the space
of linear operators on the specified Hilbert space. We can embed the
Hilbert space vectors as projectors within this larger space and
inherit the inner product in the form of the trace of an operator
product. Again, for an infinite dimensional state space this trace
must exist. As such, one is limited to trace class linear operators.

The dynamic evolution of the universe acts in these state spaces with
a time dependent unitary transform.\footnote{The generator of this transform
is possibly time-independent.} The transform is one-sided for
the Hilbert space and two-sided\footnote{As follows from the embedding.}
 for the operator space. This is also true for all other unitary
 (symmetry) transforms.
\begin{align}
\ket{\phi(t)} & =U(t_{0},t)\ket{\phi(t_{0})}\\
\Phi(t) & =U(t_{0},t)\Phi(t_{0})U(t_{0},t)^{\dagger}
\end{align}

Following postulate 4, only the sequence of states generated by the
dynamical evolution is of physical relevance. As a result, state descriptions
that are equivalent by a bijection $f$, which commutes with the evolution,
are physically indistinguishable, because it allows switching back
and forth between two representations at any point in time.

An important bijection is the left multiplication of $\Phi(t)$ with
a unitary, possibly time dependent, but easily predictable operator, that
commutes with the evolution of the universe. This operation can be
undone at any point in time and satisfies the general equivalence condition
\begin{align}
U(t_{0},t)f(\Phi)U^{\dagger}(t_{0},t) & =f(U(t_{0},t)\Phi U^{\dagger}(t_{0},t))
\end{align}
with $f(\Phi):=S\Phi$ and $[S,U(t_0,t)]=0$ for this case.
So this $f(\Phi)$ is a redundant representation of $\Phi$. However, the
redundancy does not correspond to an actual symmetry of the universe,
because those symmetries act on both sides of the state representation.
Therefore, these redundant states are not significant for the description
of the universe and have to be removed for a unique state description.
The subspace of \emph{Hermitian} trace class operators does not allow for
this one sided transform and using it for the state space gets rid of some of 
the redundant states. The \emph{non-negative} Hermitian trace class operators $\mathcal{P}$
are even exactly the \emph{quotient} of the trace class operators and equivalence
by left (or right) unitary multiplication. Hence they are an even less redundant
choice for the state space.

Both possible state spaces $\mathcal{H}$ and $\mathcal{P}$ can be
reduced even further. The linear structure of the spaces allows for
scalar multiplication as bijection that trivially commutes with the
dynamics. The two equivalence relations
\begin{align}
(\ket{a},\ket{b})\in R_{1} & \Leftrightarrow\exists c\neq0\in\mathbb{C}:\ket{a}=c\ket{b}\\
(A,B)\in R_{2} & \Leftrightarrow\exists r>0:A=rB
\end{align}
generate the quotient spaces $\mathcal{H}/R_{1}$ and $\mathcal{P}/R_{2}$
of \emph{normalized} states vectors in the Hilbert space $\mathcal{H}$ and the
non-negative \emph{unit-trace} Hermitian operator space $\mathcal{P}$ respectively.

The state space $\mathcal{P}/R_{2}$ is usually constructed as the
space of ``classical ensembles'' of quantum systems, realized by
the convex sum of projectors on the Hilbert space $\mathcal{H}$.
It is also used to describe the state of subsystems by tracing over
the environment. The first of these two uses relies strongly on the
prior existence of the measurement postulate\footnote{The construction
uses the statistics of quantum measurement and combines them with classical
ensemble statistics. This results in a compact ensemble representation,
but is not generally valid without the a-priori assumption of the
measurement postulate. Instead, one would have to list all quantum
states in the ensemble together with the probability of finding them.}
 and cannot be used for our purpose.
As for the second, tracing over the environment is not a possible
operation when describing the whole universe. And tracing is not
clearly motivated regarding the exact meaning of the result,
apart from the compatibility with the measurement postulate.
Here, the state space $\mathcal{P}/R_{2}$ was specifically constructed
from basic requirements to show that it is a valid state space, just 
like the projective Hilbert space $\mathcal{H}/R_{1}$, without the need
for an additional interpretation as space of ensembles or subsystem
states. Any meaningful interpretation must naturally follow from the behavior 
of the state representation under the system evolution. This is an important 
insight for the arguments used further down.

We can apply postulate 4 again to group dynamically indistinguishable states,
that describe the same physical behavior.
Consider a state $\Phi(t_{0})\in\mathcal{P}$ and its evolution:
\begin{align}
\Phi(t) & =U(t_{0},t)\Phi U(t_{\text{0}},t)^{\dagger}
\end{align}
Then, with the unitarity of $U$ it follows, that
\begin{align}
\Phi(t)^{2} & =U(t_{0},t)\Phi(t_{0})U(t_{0},t)^{\dagger}U(t_{0},t)\Phi(t_{0})U(t_{0},t)^{\dagger}\\
 & =U(t_{0},t)\Phi(t_{0})^{2}U(t_{0},t)^{\dagger}
\end{align}
 implying $\Phi^{2}$ evolves in exactly the same way as $\Phi$.
At any point in time, we can switch back and forth between these two states,
without making a difference dynamically, because squaring is a bijection
on non-negative Hermitian operators and commutes with the evolution.
The two states $\Phi$ and $\Phi^{2}$ are dynamically equivalent,
and as such both describe the same physical system. The same holds true
for all natural powers of a state $\Phi$, they too are bijections.

More generally, let
\begin{align}
g\,:\,\mathbb{R}_{\ge0} & \longrightarrow\mathbb{R}_{\ge0}
\end{align}
 be an analytic bijection. Such a function is necessarily strictly
monotonic and $g(0)=0$. It can be analytically extended to non-negative
Hermitian operators and is a bijection there too. Any such $g$ creates
a state, that is dynamically equivalent to the state it acts on. The
equivalence classes generated by these bijections $g$ are again a better
description of physical states.

One subset of the equivalence classes can be directly identified: 
The projectors in $\mathcal{P}$ are invariant under $g$, up to scalar 
multiplication, and each forms an equivalence class of its own. When 
$g$ acts on a state $\Phi$ it preserves two properties: The mutually
orthogonal eigensubspaces are invariant. And the ordering of the eigenvalues
is also strictly preserved, because $g$ is strictly increasing on the
non-negative real numbers. The number of eigenvalues is countable
because $\mathcal{H}$ is separable, and the trace-class bounds the
sum of the non-negative eigenvalues, so that the maximum of the eigenvalues
exists. It follows, that a possible representation of the equivalence
classes of physical states then consists of a \emph{list of orthogonal
eigensubspaces} in order of decreasing eigenvalues, and nothing else.
Particularly, the eigenvalues do not have to be specified, as they
do change with $g$. For a finite list, the last entry is the nullspace,
which is invariant under $g$. It is redundant, because it is the unique
orthogonal complement of the direct sum of the other list entries.
Therefore, this last entry may be omitted.

Depending on the equivalence class, the subspace list can have a different
number of entries. For a simple projector it only has one entry, the
eigensubspace with the maximum eigenvalue, which trivially equals 1.
That is also the one property, that all equivalence classes share.
All equivalence classes can be partially characterized with the eigensubspace
of the \emph{greatest} eigenvalue, which \emph{always} exists.

Applying the unitary evolution of the universe to the eigensubspace
lists shows, that the evolution acts independently on each subspace
and does not change the dimensions of the subspaces or the length
of the list. The character of the equivalence classes is therefore
invariant under time evolution. This effectively splits up the state
space into an infinite set of state spaces with common character and
each connected by unitary transforms.

With all these considerations in place, the result is an infinite
number of possible state spaces, that each allow for unitary evolution of
the universe and do not encode the identified symmetries.
Each state space is fully characterized by a finite list or infinite 
sequence of natural numbers or countable infinity, describing the dimensions of
a the subspaces. The evolution acts unitarily on each subspace.
For example, the original Hilbert space $\mathcal{H}$ is contained
in the set of state spaces as the sequence $(\,1\,)$, representing
a single one dimensional subspace, which the evolution acts on.

The n-dimensional projectors state space $(\, n\,)$ creates a physical
universe that slightly differs from $\mathcal{H}$. One can decide
to describe such a universe with a single vector from the eigensubspace
and will get consistent dynamics. So the one dimensional state description
is not unique, and even several different one dimensional descriptions could
be used at the same time. This universe can be understood as a generated
by $n$ non-interfering superpositions of orthogonal, but not uniquely
determined $\mathcal{H}$ universes. Observers would notice the absence
of interference in certain situations. But for finite $n$ that
effect would be so rare that such a universe is practically indistinguishable
from a $(\,1\,)$ universe. Furthermore a coincidence of eigenvalues
to form a more than 1-dimensional eigensubspace is very unlikely,
in fact those configurations form a subset of measure zero for all
practical measures on the operator representation.

For universes with a state space with more than one relevant subspace
$(\, n,\ldots\,)$, the description of their behavior can be simplified further.
There must be, at least in principle, an experiment, that can distinguish different 
values of a quantity for it to be of physical relevance. 
This is equivalent to saying, the evolution must be able to produce different
outcomes depending on that quantity. In the case of the unitary evolution we 
are discussing here, there is no way for the dynamics to depend on the \emph{number}
of subspaces in the state list. The number of subspaces is therefore not a feature
of the state's behavior. It is not physically relevant.

With the number of subspaces inherently unknown, the description of the individual
subspaces also cannot be relevant. Consequently, there is not much left to be used 
for a state description. We know from deduction, that one subspace must always exist;
the one with the \emph{greatest eigenvalue}.

This can also be motivated by applying an explicit bijection 
\begin{align}
g_{k}(\Phi) & =\Phi^{k}\label{eq:PowerLimit}
\end{align}
for natural $k$ to a non-negative trace class Hermitian operator representation.
For sufficiently large $k$, the largest eigenvalue dominates as strongly
as desired. So the dynamics of the equivalence class representatives
realized as a state operator can get arbitrarily close to the dynamics
of the states in $(\, n\,)$. And for $k\to\infty$ we recover that
space as the limit. As a result, there is no new physics to be
expected in $(\, n,\ldots\,)$ universes compared to an $(\, n\,)$
universe, which is an \emph{accumulation point} of the equivalence class. 

It follows, that the map $g_{k}$ for $k\to\infty$, corresponding
to the projection onto the eigensubspace with the largest eigenvalue,
is a useful tool for reducing the complexity of the dynamical description
in non-projection state space universes.

Summing up, the behavior of a unitarily evolving system is fully captured
in a $(\,n\,)$ state description, which for $n=1$ is reduced to the 
traditional Hilbert space $\mathcal{H}$. This conclusion and the 
corresponding construction will be used further down to describe the
behavior of the universe, as perceived by a local observer. There, additional
constraints due to the light-cone structure of the universe will become relevant.

\subsection{Local interactions and available state information}

One of the most obvious weaknesses of the measurement postulate of
traditional quantum theory lies in the lack of a precise definition
of an observer. For our purpose, an observer must be a mechanism%
\footnote{A \emph{mechanism} relies on interactions and thus the
Einstein-local evolution by definition. So it is automatically localized.%
}
equipped with a memory%
\footnote{The memory requirement is somewhat optional. It guarantees
that the observer can keep track of outcomes and talk about outcome statistics.%
}
and realized by physical interactions in the universe.
It analyzes\footnote{This statement is totally agnostic of the
 strategy applied for analysis.} its environment, stores information
and makes predictions based on these observations. This comes with the
implicit emergence of a certain subjective reality for the observer, 
based on the behavior\footnote{Read: interactions between the observer 
and the universe.} of the universe around him.

An observer of this kind can be human\footnote{Ideally a physicist, of course!},
but it does not have to be. The complexity and details of the mechanism are
not of importance for the essential properties of our observer.
Accordingly, these aspects will not be part of the model. The mere
presence of an observer within the system will be assumed, even if
the number of degrees of freedom is much too small to actually contain any
sensible realization of such an observer mechanism.%
\footnote{This is not really a limitation of the theory. We will simply be using
the same information gathering methods, that are available to a hypothetical
observer, without being interested in its actual physical details.
As we will see, this approach is entirely sufficient.}
We will also assume, that there is a rest frame for the observer.
A discussion of the precise nature of the observer-centric constraints
follows.

\subsubsection{Stripping the universe}

In its rest frame, we can assume the relevant parts of the observer
to be contained in a spherical spatial region of radius $r_{l}$.
The universe outside of this radius is not part of the observer mechanism,
but it does influence the observer by interaction up to a radius 
\begin{align}
r_{h}: & =r_{l}+c\, T
\end{align}
limited by the observation time $T$ of the observer and the upper speed
bound $c$ for the propagation of the effect of interactions. During the
observation, this radius is a strict horizon for the information, the observer 
can recover about the universe. Anything outside the horizon is 
dynamically inaccessible and therefore not part of the local behavior 
and his experience of reality.

In a \emph{classical} universe, we could simply remove all objects
or fields on the far side of the horizon and the observer would not
notice. The observer can only reconstruct the state of the universe 
as one without any structure outside the horizon, if he does not want 
to rely on arbitrary guesswork. 

In principle, the same applies to a \emph{quantum} universe, but with
a non-trivial complication. Each particle carries its own copy of spatial
information, instead of sharing one spatial background like in a classical
universe. Starting with a single particle Hilbert space $\mathcal{H}_{1}$
and ignoring indistinguishability and different kinds of particles for now,
we can define the $n$-particle space by taking the tensor product power 
\begin{align}
\mathcal{H}_{n}: & =\mathcal{H}_{1}^{\otimes n}
\end{align}
and in the next step the Fock space of the particles as the direct
sum of all possible $n$-particle spaces. The one dimensional vacuum
space is denoted by $\mathcal{H}_{1}^{\otimes 0}$. 
\begin{align}
\mathcal{H}_{F} & :=\bigoplus_{n=0}^{\infty}\mathcal{H}_{1}^{\otimes n}
\end{align}
For a single particle space, we can introduce two projection operators
$P_{a}$ and $P_{i}$, that project onto the dynamically accessible
and inaccessible states respectively. Naturally, $P_{a}+P_{i}=\mathbf{1}$
must hold. In case that accessibility is determined by the horizon
with radius $r_{h}$ as defined above, we can explicitly express the
projectors in the position basis of the single particle Hilbert space.
\begin{align}
P_{a} & :=\int_{r\le r_{h}}\ket{\mathbf{r}}\bra{\mathbf{r}}d^{3}\mathbf{r}\\
P_{i} & :=\int_{r>r_{h}}\ket{\mathbf{r}}\bra{\mathbf{r}}d^{3}\mathbf{r}
\end{align}

The eigensubspaces of both of these projectors, $\mathcal{H}_{a}$
and $\mathcal{H}_{i}$, hold the accessible and inaccessible states.
We can write the single particle Hilbert space as a direct sum of
these two orthogonal subspaces. 
\begin{equation}
\mathcal{H}_{1}=\mathcal{H}_{a}\oplus\mathcal{H}_{i}
\end{equation}

Expanding the Fock space definition with this substitution results
in
\begin{align}
\mathcal{H}_{F} & =\mathcal{H}_{1}^{\otimes0}\oplus\bigoplus_{n=1}^{\infty}\mathcal{H}_{a}^{\otimes n}\oplus\bigoplus_{n=1}^{\infty}\mathcal{H}_{i}^{\otimes n}\label{eq:Fock Space Splitting}\\
 & \oplus\bigoplus_{n,m=1}^{\infty}S\left(\mathcal{H}_{a}^{\otimes n},\mathcal{H}_{i}^{\otimes m}\right)\nonumber 
\end{align}
where $S\left(\mathcal{H}_{a}^{\otimes n},\mathcal{H}_{i}^{\otimes m}\right)$
is the symmetrized direct sum of the tensor product of $n$ times
$\mathcal{H}_{a}$ and $m$ times $\mathcal{H}_{i}$. For example
the symmetrization $S\left(\mathcal{H}_{a}^{\otimes2},\mathcal{H}_{i}^{\otimes1}\right)$
expands to: 
\begin{align}
S\left(\mathcal{H}_{a}^{\otimes2},\mathcal{H}_{i}^{\otimes1}\right)= & \left(\mathcal{H}_{a}\otimes\mathcal{H}_{a}\otimes\mathcal{H}_{i}\right)\nonumber \\
\oplus & \left(\mathcal{H}_{a}\otimes\mathcal{H}_{i}\otimes\mathcal{H}_{a}\right)\nonumber \\
\oplus & \left(\mathcal{H}_{i}\otimes\mathcal{H}_{a}\otimes\mathcal{H}_{a}\right)
\end{align}
If we want to strip the inaccessible information from $\mathcal{H}_{F}$
in equation \eqref{eq:Fock Space Splitting}, we can first drop%
\footnote{As will be discussed, they actually have to be replaced
with a vacuum state.%
} the pure $\mathcal{H}_{i}$ powers, similarly to the classical situation.
The remaining inaccessible parts of the state space appear in tensor
products with the accessible subspace. Removing these inaccessible
parts, while keeping as much as possible of the accessible parts intact,
will also remove relative phase information contained in the eliminated 
factor. This is in full agreement with the preservation of the local behavior.
If we have two state components, that are orthogonal in the full state space,
then they will not interfere. The reduction procedure may however map 
them to identical local states, which are still not allowed to interfere,
because that is the behavior to be preserved. 

So we cannot add the two states coherently. We would not know the relative 
phase anyway. The stripped state space must therefore contain a way to superimpose
states without interference, while containing the Fock space 
$\mathcal{H}_{F}$. 

The expanded state space for local reductions is given by the trace class 
non-negative hermitian operators on $\mathcal{H}_{F}$ denoted by 
$T(\mathcal{H}_{F})$, with the canonical embedding of the first as projections 
in the latter. The dynamics preserving map
\begin{align}
S\left(\mathcal{H}_{a}^{\otimes n},\mathcal{H}_{i}^{\otimes m}\right) & \to T\left(\mathcal{H}_{a}^{\otimes n}\right)
\end{align}
can then be realized by tracing over the inaccessible tensor factor
spaces. This operation keeps the relative weights and eliminates
all structural information in the inaccesible factor spaces without
touching the accessible factor spaces.

Let $\ket{\Psi}$ be a Fock state and $\ket{\psi_{n}}$ its $n$-particle
component.
\begin{align}
\ket{\Psi} & =\sum_{n=0}^{\infty}c_{n}\ket{\psi_{n}}
\end{align}
The projection operators $P_{a}^{(k,n)}$ and $P_{i}^{(k,n)}$ act
on the $k$-th particle in the $n$-particle subspace. Stripping only
the 1-particle subspace starts with distributing the projectors over
the state.
\begin{align}
\ket{\psi_{1}} & =\left(P_{a}^{(1,1)}+P_{i}^{(1,1)}\right)\ket{\psi_{1}}\\
 & =P_{a}^{(1,1)}\ket{\psi_{1}}+P_{i}^{(1,1)}\ket{\psi_{1}}\nonumber 
\end{align}
If the particle is entirely accessible and therefore 
\begin{align}
P_{i}^{(1,1)}\ket{\psi_{1}} & =0
\end{align}
 we already have a stripped state. If the particle is entirely inaccessible
we have 
\begin{align}
P_{a}^{(1,1)}\ket{\psi_{1}} & =0
\end{align}
and dropping the inaccessible part would result in a physically undefined
state, the $0$-vector. Instead, the observer should simply find no
particle, or in other words, the vacuum state $\ket{\psi_{0}}$. The
vacuum state can already be part of $\ket{\Psi}$ however, and the
observed subjective vacuum must not interfere with the true vacuum.%
\footnote{This is an example of how in the full unreduced space the states 
do not interfere, because they are orthogonal, but are then mapped to identical
local states. So to preserve the behavior, their images cannot interfere
in the reduced space.%
}
The two vacua must add incoherently. To simplify the notation we can
identify the vacuum space $\mathcal{H}_{1}^{\otimes0}$ with the complex
scalars and choose: 
\begin{align}
\ket{\psi_{0}} & =1\in\mathbb{C}
\end{align}
So if we apply the stripping map 
\begin{align}
\Lambda:\mathcal{H}_{F} & \to T\left(\mathcal{H}_{F}\right)
\end{align}
to the vacuum and single particle subspaces we get
\begin{align}
 & \Lambda\left(c_{0}\ket{\psi_{0}}+c_{1}\ket{\psi_{1}}\right)\\
 & =\Lambda\left(c_{0}\ket{\psi_{0}}+c_{1}P_{a}^{(1,1)}\ket{\psi_{1}}+c_{1}P_{i}^{(1,1)}\ket{\psi_{1}}\right)\nonumber \\
 & =T\left(c_{0}\ket{\psi_{0}}+c_{1}P_{a}^{(1,1)}\ket{\psi_{1}}\right)+\mathrm{tr}_{\mathcal{H}_{1}}\left(T\left(c_{1}P_{i}^{(1,1)}\ket{\psi_{1}}\right)\right)\nonumber 
\end{align}
where 
\begin{align}
T\left(\ket{\phi}\right): & =\ket{\phi}\bra{\phi}
\end{align}
is the natural embedding. The trace results in a complex number representing
the vacuum state. 

If we also consider two-particle states, we get additional terms from
\begin{align}
\ket{\psi_{2}} & =\left(P_{a}^{(1,2)}+P_{i}^{(1,2)}\right)\left(P_{a}^{(2,2)}+P_{i}^{(2,2)}\right)\ket{\psi_{2}}\\
 & =P_{a}^{(1,2)}P_{a}^{(2,2)}\ket{\psi_{2}}+P_{i}^{(1,2)}P_{i}^{(2,2)}\ket{\psi_{2}}\nonumber \\
 & +P_{a}^{(1,2)}P_{i}^{(2,2)}\ket{\psi_{2}}+P_{i}^{(1,2)}P_{a}^{(2,2)}\ket{\psi_{2}}\nonumber 
\end{align}
The first term is not changed by $\Lambda$. The second term is mapped
to the vacuum. And the last two terms are traced over to result in
an incoherently mixing single particle state.
\begin{align}
 & \Lambda\left(c_{0}\ket{\psi_{0}}+c_{1}\ket{\psi_{1}}+c_{2}\ket{\psi_{2}}\right)\\
 & =T\left(c_{0}\ket{\psi_{0}}+c_{1}P_{a}^{(1,1)}\ket{\psi_{1}}+c_{2}P_{a}^{(1,2)}P_{a}^{(2,2)}\ket{\psi_{2}}\right)\nonumber \\
 & +\mathrm{tr}_{\mathcal{H}_{i}}\left(T\left(c_{1}P_{i}^{(1,1)}\ket{\psi_{1}}+c_{2}P_{a}^{(1,2)}P_{i}^{(2,2)}\ket{\psi_{2}}\right.\right.\nonumber \\
 & \quad\left.\left.+c_{2}P_{i}^{(1,2)}P_{a}^{(2,2)}\ket{\psi_{2}}\right)\right)\nonumber \\
 & +\mathrm{tr}_{\mathcal{H}_{i}^{\otimes2}}\left(T\left(c_{2}P_{i}^{(1,2)}P_{i}^{(2,2)}\ket{\psi_{2}}\right)\right)\nonumber 
\end{align}
The trace is over all terms with a single inaccessible particle. The
different position of the traced tensor factor space in each term is not problematic.
Reordering the inaccessible spaces does not change the result, as they
are traced over eventually. So we can think of that reordering implicitly
happening while tracing.

As can be seen, this becomes tedious and complicated to write down
for an increasing number of particle number eigenspace contributions.
The definition of $\Lambda$ is a lot easier, if we express it in terms
of an occupation number basis of $\mathcal{H}_{F}$, starting from a single
particle eigenbasis of the projection operators $P_{a}$ and $P_{i}$ and assuming
indistinguishable particles this time. With the bases $\ket{k}_{a}$
and $\ket{k}_{i}$ for $\mathcal{H}_{a}$ and $\mathcal{H}_{i}$ respectively
and natural $k$, we can characterize a fully symmetric $n$-particle
state by listing the number of particles $n_{k;a}$ and $n_{k;i}$
in each accessible and inaccessible single particle base state.
The total particle number is then
\begin{align}
n & =\sum_{k=0}^{\infty}\left(n_{k;a}+n_{k;i}\right)
\end{align}
and we can label the state
\begin{equation}
\ket{\left(n_{0;a},n_{1;a},n_{2;a},\cdots\right);\left(n_{0;i},n_{1;i},n_{2;i},\cdots\right)}
\end{equation}
or in a more compact form
\begin{equation}
\ket{\mathbf{n}_{a};\mathbf{n}_{i}}
\end{equation}
with the occupation lists $\mathbf{n}_{a}$ and $\mathbf{n}_{i}$
respectively.

For our convenience we define
\begin{align}
\left|\mathbf{n}_{a}\right|: & =\sum_{k=0}^{\infty}n_{k;a}
\end{align}
and likewise for $\left|\mathbf{n}_{i}\right|$, so that: 
\begin{align}
n & =\left|\mathbf{n}_{a}\right|+\left|\mathbf{n}_{i}\right|
\end{align}

With the occupation number basis of the Fock space, it can be seen
from direct calculation, that $\ket{\mathbf{n}_{a};0}$ remains unchanged
under $\Lambda$, while states with inaccessible single particle states
are mapped to stripped states. 
\begin{align}
\ket{\mathbf{n}_{a};\mathbf{n}_{i}} & \mapsto T\left(\ket{\mathbf{n}_{a};0}\right)
\end{align}
This already includes the case of all inaccessible states
being mapped to the vacuum. The full stripping map is then defined
for both bosonic and fermionic states and explicitly given by the
following expression.
\begin{align}
\Lambda\,:\,\mathcal{H}_{F} & \longrightarrow T\left(\mathcal{H}_{F}\right)\\
\ket{\Psi} & \longmapsto\Lambda\left(\ket{\Psi}\right)\nonumber \\
\Lambda\left(\ket{\Psi}\right) & :=\sum_{\mathbf{n}_{i}}T\left(\sum_{\mathbf{n}_{a}}\ket{\mathbf{n}_{a};0}\braket{\mathbf{n}_{a};\mathbf{n}_{i}|\Psi}\right)\nonumber 
\end{align}

For an entirely \emph{accessible} state we get
\begin{align}
\Lambda\left(\ket{\Psi_a}\right) & =\ket{\Psi_a}\bra{\Psi_a}
\end{align}
while fully \emph{inaccessible} states are mapped to the vacuum.
\begin{align}
\Lambda\left(\ket{\Psi_i}\right) & =\ket{\circ}\bra{\circ}
\end{align}

The construction of the stripping map guarantees, that an observer
interacting with the universe will not be able to distinguish the
real state of the universe $\ket{\Psi}$ from the stripped state $\Lambda(\ket{\Psi})$.
He certainly cannot construct a more accurate description of the universe
than the stripped state. But does he even have the means to reconstruct the stripped
state itself? The answer must be, that he cannot, for the following two reasons.

First, the classification of accessible and inaccessible states based
on the horizon radius discussed above is not very precise. While all
states marked as dynamically inaccessible are in fact certainly not
accessible, not all states marked as accessible really are practically accessible.
That means, the stripped state contains information, that is not available
to the observer even under almost ideal conditions. We will discuss this
in more depth below, when we look at the process of observing an experiment.

The second reason is the inability of the observer to distinguish
state representations, that are equivalent under unitary evolution.
The localized state description will not evolve unitarily in general,
nonetheless unitarity is an important special case for which the
reduction has to be \emph{applicable} and \emph{consistent}.
Like discussed in section \ref{sec:StatUni}, $\Lambda(\ket{\Psi})$
is, for example, intrinsically indistinguishable from its positive integer
powers, if we assume unitary evolution. Consequently, the observer does 
not have enough information to reconstruct the stripped state without 
further assumptions about the state of the universe. As we have seen,
the \emph{only} assumption, that does not require the addition of arbitrarily
made up information is, that the reconstructed state of the universe is a 
$(\,n\,)$ state. Following the earlier discussion about the state of the 
universe, we define the normalized stripped state in $\mathcal{H}_{F}$ as
the normalized limit of the bijection \eqref{eq:PowerLimit}.
\begin{align}
\bar{\Lambda}:\mathcal{H}_{F}^{\prime} & \longrightarrow\mathcal{H}_{F}\\
\ket{\Psi} & \longmapsto T^{-1}\left(\bar{\Lambda}(\ket{\Psi})\right)\nonumber \\
\bar{\Lambda}(\ket{\psi}) & :=\lim_{k\to\infty}\frac{\Lambda(\ket{\Psi})^{k}}{\mathrm{tr}\left(\Lambda(\ket{\Psi})^{k}\right)}\nonumber 
\end{align}

The limit is a projection operator onto the eigensubspace of $\Lambda(\ket{\Psi})$
with the greatest eigenvalue. For a so called pure state, the eigensubspace must
be one dimensional, which can then be naturally%
\footnote{This mapping is not unique, because of the arbitrary phase. All possible
maps work equally well however, because the chosen phase is global
and commutes with the evolution.%
} mapped back to the Hilbert space $\mathcal{H}_{F}$ by $T^{-1}$.
If we assume, that the eigenvalues are essentially random, then it is
extremely unlikely, that eigenvalues become exactly equal and form an
eigensubspace with dimension greater than one.
So it seems to be safe to assume, that $\bar{\Lambda}(\ket{\Psi})$
practically always results in a pure state. The states that do not
get mapped back to a pure state are formally removed from $\mathcal{H}_{F}^{\prime}$.
We will see, that the time evolution of the resulting state becomes essentially
non-linear once we drop the unitarity assumption for the \emph{stripped} state evolution.

\subsubsection{Making an observation}

The process of observation is based on interactions and limited
by their finite propagation speed in a relativistic universe. Hence,
the reconstructed state refers to a time slice in the \emph{past} of the
observer. This time asymmetry is created by the requirement of a memory
to consolidate the state information. The implied arrow of time does not
affect the measurement process at this point and reconstructing
a future state leads to the same fundamental properties. We will see, that
the process itself will allow for an independent time arrow to be inherited
from the radiation field.

\begin{figure}
\begin{centering}
$ $\includegraphics[width=1\columnwidth]{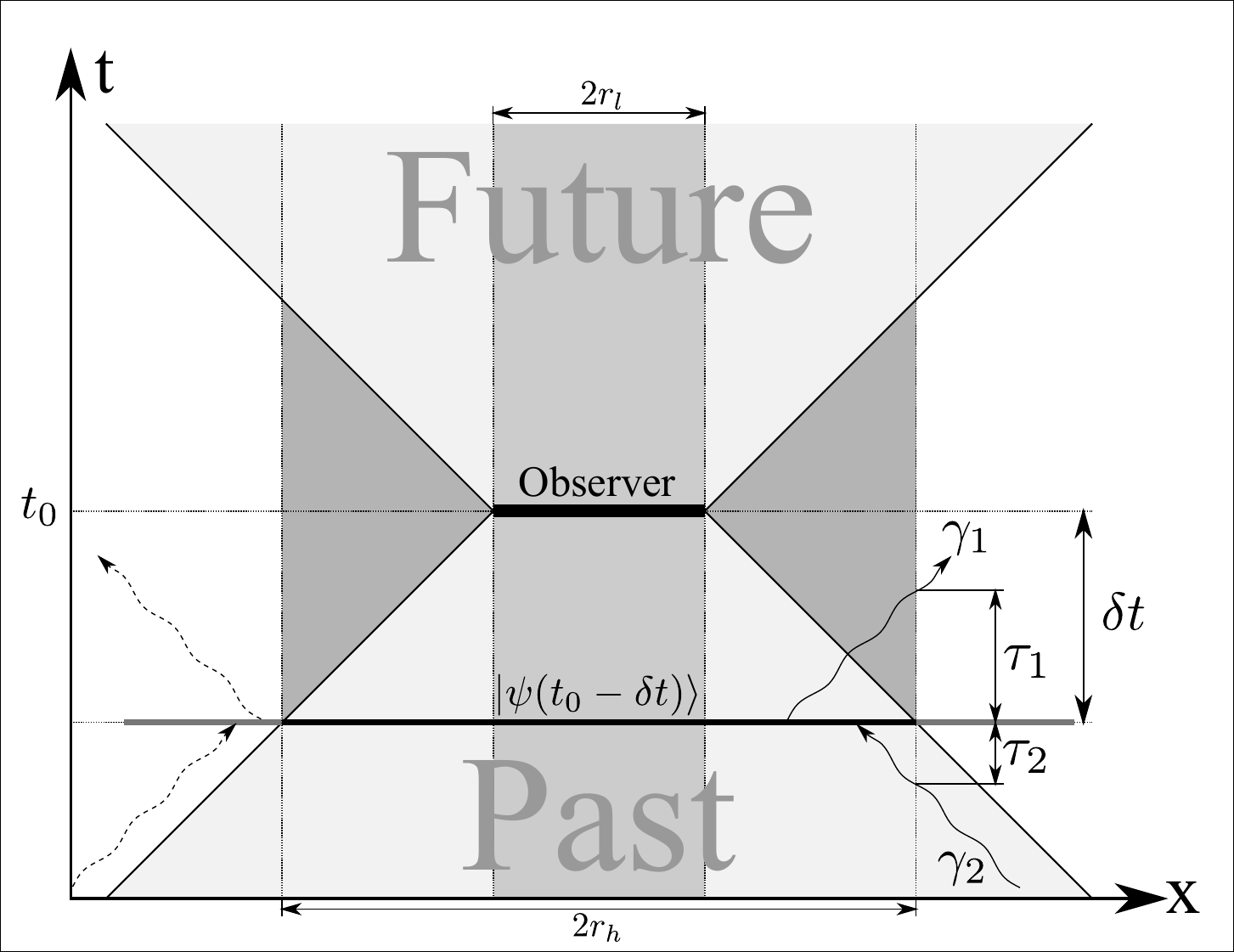}
\par\end{centering}

\caption{\label{fig:ObserverReconstruction}The observer reconstructs the state
$\ket{\psi}$ based on the information collected from past interactions.
The photons $\gamma_{1}$ and $\gamma_{2}$ transport information
to and from the environment beyond the observation horizon.}
\end{figure}

Figure \ref{fig:ObserverReconstruction} shows the situation in spacetime.
The observer with radius $r_{l}$ reconstructs the state $\ket{\psi(t_{0}-\delta t)}$
at the time $t_{0}$ from information in his past light cone. The
observation delay $\delta t$ and the radius of the reconstructed
state horizon are related by $r_{h}=r_{l}+c\,\delta t$. During an
observation the delay is held constant, so that the observed system
evolves on the same time base as the observer. Anything outside the
world volume swept by the space inside the horizon is not part of
the observer's subjective reality.

In a typical experimental setup, the observer focuses on a more or
less isolated object at relative rest. It is a reasonable simplification
to assume, that the only exchange of information with the unobserved
universe is due to electromagnetic interactions. The two photons shown
on the left do not have any effect on the reconstructed reality
of the observer, because their interaction is constrained to the spatial
region outside the reconstruction horizon. This is not true for the
photons $\gamma_{1}$ and $\gamma_{2}$. They do interact with the reconstructed
state. The corresponding durations of the interaction with the world
volume of the state are $\tau_{1}$ and $\tau_{2}$. During that time,
the state reconstruction performed by the observer will change in
a non-unitary way. Furthermore, the observer will not be able to directly
reconstruct the existence of either photon, because it does not pass
through his own world volume.

As discussed before, the reconstructed state is described by a single
Hilbert space vector. The outgoing and incoming photons are of unknown
polarization state, which leads to a random element being introduced to
the reconstructed state. The change of the state happens on the same
, almost instant, time scale as the typical interaction with a photon.
The reconstructed subjective state description will therefore change
suddenly and non-unitarily to a random outcome state. This is not only
in agreement with the measurement postulate. At this point it seems natural
to conjecture a relationship between the interaction with photons and what
is called quantum measurement. The next chapter discusses the 
relevant process in more detail.

\section{Simple interactions between Radiation and Matter}

Let us consider a system consisting of only a single qubit. There
is nothing else, the observer can use to compare the qubit to. Not
even the observer itself exists as part of the quantum system in this
simple model. The qubit is the entire observable universe, and
we can apply the reconstruction procedure to it. 

Nonetheless, the qubit is not completely insulated, as we assume it couples 
to the radiation field. The interaction is described by a unitary evolution
on the entire unobserved universe. To keep it simple, the only states of
the radiation field, that are considered, are inaccessible. The incoming
state can be thought of as a single photon with a 2-dimensional representation
of its polarization. The outgoing states can more complicated than
that, depending on the scattering process, and we do not specifically keep track
of polarization. We call these scattering processes \emph{elementary}, because only two 
qubits, one for the photon and one local to the observer, are given as the input state.

We will discuss three canonical cases of elementary scattering processes
explicitly. In all of them, the state of the local qubit is represented in the
orthonormal basis $\{\ket{0},\ket{1}\}$, which is chosen to coincide with
the \emph{physically} preferred basis\footnote{We do not postulate an  a-priori
preferred basis. We only use the knowledge of the basis preferred by the process
to simplify the presentation.} for the resulting process. The incoming photon
state will be written as a linear combination of the orthonormal basis states
$\{\ket{\leftrightarrow},\ket{\updownarrow}\}$.\footnote{Even though the
labels are quite suggestive, this does not have to 
be the basis of linear polarization states. The arrows merely symbolize any basis
of the polarization space, including circular.} For the outgoing
radiation states an arbitrary\footnote{The radiation field after
the photon collision is more complicated and will in general
require more basis vectors.} orthonormal basis 
$\{\ket{\rightsquigarrow_{n}}\,:\, n\in\mathbb{N}\}$ is used. 
The vacuum state is $\ket{\circ}$. All processes are unitary,
because they map orthonormal states in the input space to orthonormal
states in the output space.

The input state corresponds to the objective
state of the universe just before the interaction. And the output state
is the objective state just after the interaction. Both, input and
output radiation states are locally inaccessible and will be removed
in the local state reconstruction, so that we can study the scattering
process as observed locally. The state of the photon is considered
to be \emph{entirely unknown}. We will write the photon component of the input 
state to the left of the qubit component and for the output state to the 
right of the qubit, only to indicate the incoming and outgoing 
radiation and without any deeper mathematical or physical meaning.

The general%
\footnote{This is clearly not the most general state for the given input space, 
because it assumes separable photon and qubit states. We will be able to relax this 
restriction somewhat when the concatenation of scattering events will be discussed. 
For now, we have to assume an additional external arrow of time that allows the 
separation of qubit and photon by assuming no interaction between the systems in the past.%
} input state in this setup is 
\begin{align}
\ket{\psi} & :=\left(\alpha\ket{\updownarrow}+\beta\ket{\leftrightarrow}\right)\left(a\ket{0}+b\ket{1}\right)\\
 & =\alpha a\ket{\updownarrow}\ket{0}+\alpha b\ket{\updownarrow}\ket{1}\nonumber \\
 & +\beta a\ket{\leftrightarrow}\ket{0}+\beta b\ket{\leftrightarrow}\ket{1}
\end{align}
for $\alpha,\beta,a,b\,\in\mathbb{C}$ with the non-vanishing condition: 
\begin{align}
\alpha\alpha^{*}+\beta\beta^{*} & >0\\
aa^{*}+bb^{*} & >0
\end{align}
The local reconstruction of the input $\ket{\psi}$ is then simply 
\begin{align}
\bar{\Lambda}(\ket{\psi}) & =T((a\ket{0}+b\ket{0})\ket{\circ})
\end{align}
which is equivalent to the local qubit state, just like one would expect.

\subsection{The uniform elementary scattering process}

Consider the following unitary mapping of the input space basis to the output space
basis:

\begin{align}
U_{u}\,:\, & \left.\begin{array}{r}
\ket{\updownarrow}\ket{\text{0}}\\
\ket{\updownarrow}\ket{1}\\
\ket{\leftrightarrow}\ket{0}\\
\ket{\leftrightarrow}\ket{1}
\end{array}\right\} \longmapsto\left\{ \begin{array}{l}
\ket{0}\ket{\rightsquigarrow_{1}}\\
\ket{0}\ket{\rightsquigarrow_{2}}\\
\ket{1}\ket{\rightsquigarrow_{3}}\\
\ket{1}\ket{\rightsquigarrow_{4}}
\end{array}\right.
\end{align}
The objective input state $\ket{\psi}$ is then mapped to
\begin{align}
U_{u}\ket{\psi} & =\alpha a\ket{0}\ket{\rightsquigarrow_{1}}+\alpha b\ket{0}\ket{\rightsquigarrow_{2}}\\
 & +\beta a\ket{1}\ket{\rightsquigarrow_{3}}+\beta b\ket{1}\ket{\rightsquigarrow_{4}}\nonumber 
\end{align}
The stripped local state of the output
is then
\begin{align}
\Lambda(U_{u}\ket{\psi}) & =T\left(\alpha a\ket{0}\ket{\circ}\right)+T\left(\alpha b\ket{0}\ket{\circ}\right)\\
 & +T\left(\beta a\ket{1}\ket{\circ}\right)+T\left(\beta b\ket{1}\ket{\circ}\right)\nonumber \\
 & =\alpha\alpha^{*}(aa^{*}+bb^{*})\ket{0}\bra{0}\ket{\circ}\bra{\circ}\\
 & +\beta\beta^{*}(aa^{*}+bb^{*})\ket{1}\bra{1}\ket{\circ}\bra{\circ}\nonumber 
\end{align}
After taking the projection limit and mapping the result into the Hilbert space, we get
\begin{align}
\bar{\Lambda}(U_{u}\ket{\psi}) & =\begin{cases}
\ket{0}\ket{\circ} & \text{if}\;\alpha\alpha^{*}>\beta\beta^{*}\\
\ket{1}\ket{\circ} & \text{if}\;\alpha\alpha^{*}<\beta\beta^{*}
\end{cases}
\end{align}
The state of the photon is entirely unknown and we can assume that
any polarization state has the same probability.\footnote{To be
very clear, the photon carries one specific polarization state.
We just do not know which one is actually realized. There is no
involvement of the measurement postulate, any ad-hoc probability
measure, or even density matrices.} Then the symmetry of the
result allows the deduction of the probabilities of either outcome
and we find both to be equally likely.
The case of exact equality of $\alpha\alpha^*$ and $\beta\beta^*$
is of zero measure and can be ignored in the statistics.

Summarizing, we have a process $U$ that acts locally like
\begin{align}
U_{u}\,:\, a\ket{0}+b\ket{1} & \mapsto\begin{cases}
\ket{0} & \text{with probability}\, p_{\text{0}}=\frac{1}{2}\\
\ket{1} & \text{with probability}\, p_{1}=\frac{1}{2}
\end{cases}
\end{align}

Notably, the result does not depend on the state of the local input qubit.
A local observer just sees a fully random outcome that can be predicted
only in terms of probabilities, while the global process is entirely
deterministic. The missing information about the exact photon state creates
the illusion of a spontaneous random change of the local quantum state.

\subsection{The maximum elementary scattering process}

After having seen a process whose result only depends on the photon state,
the existence of a process with only local qubit dependence is not very
surprising. A small, yet significant, modification of the unitary map generates
such a scattering result:
outcome:
\begin{align}
U_{m}\,:\, & \left.\begin{array}{r}
\ket{\updownarrow}\ket{\text{0}}\\
\ket{\updownarrow}\ket{1}\\
\ket{\leftrightarrow}\ket{0}\\
\ket{\leftrightarrow}\ket{1}
\end{array}\right\} \longmapsto\left\{ \begin{array}{l}
\ket{0}\ket{\rightsquigarrow_{1}}\\
\ket{1}\ket{\rightsquigarrow_{2}}\\
\ket{0}\ket{\rightsquigarrow_{3}}\\
\ket{1}\ket{\rightsquigarrow_{4}}
\end{array}\right.
\end{align}
The objective input state $\ket{\psi}$ is mapped to
\begin{align}
U_{m}\ket{\psi} & =\alpha a\ket{0}\ket{\rightsquigarrow_{1}}+\alpha b\ket{1}\ket{\rightsquigarrow_{2}}\\
 & +\beta a\ket{0}\ket{\rightsquigarrow_{3}}+\beta b\ket{1}\ket{\rightsquigarrow_{4}}\nonumber 
\end{align}
which is stripped and results in
\begin{align}
\Lambda(U_{m}\ket{\psi}) & =T\left(\alpha a\ket{0}\ket{\circ}\right)+T\left(\alpha b\ket{1}\ket{\circ}\right)\\
 & +T\left(\beta a\ket{0}\ket{\circ}\right)+T\left(\beta b\ket{1}\ket{\circ}\right)\nonumber \\
 & =aa^{*}(\alpha\alpha^{*}+\beta\beta^{*})\ket{0}\bra{0}\ket{\circ}\bra{\circ}\\
 & +bb^{*}(\alpha\alpha^{*}+\beta\beta^{*})\ket{1}\bra{1}\ket{\circ}\bra{\circ}\nonumber 
\end{align}
Eventually, with the limit and the Hilbert space map in place we arrive at
\begin{align}
\bar{\Lambda}(U_{m}\ket{\psi}) & =\begin{cases}
\ket{0}\ket{\circ} & \text{for}\, aa^{*}>bb^{*}\\
\ket{1}\ket{\circ} & \text{for}\, aa^{*}<bb^{*}
\end{cases}
\end{align}
So in total we get the local observation of the form
\begin{align}
U_{m}\,:\, a\ket{0}+b\ket{1} & \mapsto\begin{cases}
\ket{0} & \text{if}\;|a|>|b|\\
\ket{1} & \text{if}\;|b|>|a|
\end{cases}
\end{align}
The unknown photon state does not influence the outcome. And the process
is fully deterministic, even for a local observer. The result is a 
perceived projection onto the dominant component of the qubit in the 
preferred basis.

\subsection{The Born rule generating elementary scattering process}

The two scattering processes discussed above only use information
from either the qubit or the photon state to determine the local output.
Next, we will discuss a process that mixes the influence of
both equally and results in a statistical rule that also depends on
the state of the qubit. The map, that creates this behavior is
\begin{align}
U_{B}\,:\, & \left.\begin{array}{r}
\ket{\updownarrow}\ket{\text{0}}\\
\ket{\updownarrow}\ket{1}\\
\ket{\leftrightarrow}\ket{0}\\
\ket{\leftrightarrow}\ket{1}
\end{array}\right\} \longmapsto\left\{ \begin{array}{l}
\ket{0}\ket{\rightsquigarrow_{1}}\\
\left(\ket{1}\ket{\rightsquigarrow_{2}}+\ket{0}\ket{\rightsquigarrow_{3}}\right)/\sqrt{2}\\
\left(\ket{0}\ket{\rightsquigarrow_{4}}+\ket{1}\ket{\rightsquigarrow_{5}}\right)/\sqrt{2}\\
\ket{1}\ket{\rightsquigarrow_{6}}
\end{array}\right.
\end{align}
and produces the output state
\begin{align}
U_{B}\ket{\psi} & =\alpha a\ket{0}\ket{\rightsquigarrow_{1}}+\beta b\ket{1}\ket{\rightsquigarrow_{6}}\\
 & +\alpha b\left(\ket{1}\ket{\rightsquigarrow_{2}}+\ket{0}\ket{\rightsquigarrow_{3}}\right)/\sqrt{2}\nonumber \\
 & +\beta a\left(\ket{0}\ket{\rightsquigarrow_{4}}+\ket{1}\ket{\rightsquigarrow_{5}}\right)/\sqrt{2}\nonumber 
\end{align}
which strips to
\begin{multline}
\Lambda(U_{B}\ket{\psi})=T\left(\alpha a\ket{0}\ket{\circ}\right)\\
+\frac{1}{2}\left(T\left(\alpha b\ket{1}\ket{\circ}\right)+T\left(\alpha b\ket{0}\ket{\circ}\right)\right)\\
+\frac{1}{2}\left(T\left(\beta a\ket{0}\ket{\circ}\right)+T\left(\text{\ensuremath{\beta}a}\ket{1}\ket{\circ}\right)\right)\\
+T\left(\beta b\ket{1}\ket{\circ}\right)\\
=\left(\alpha\alpha^{*}aa^{*}+\frac{1}{2}\left(\alpha\alpha^{*}bb^{*}+\beta\beta^{*}aa^{*}\right)\right)\ket{0}\bra{0}\ket{\circ}\bra{\circ}\\
+\left(\beta\beta^{*}bb^{*}+\frac{1}{2}\left(\alpha\alpha^{*}bb^{*}+\beta\beta^{*}aa^{*}\right)\right)\ket{1}\bra{1}\ket{\circ}\bra{\circ}
\end{multline}
and after mapping the result to pure states we get:
\begin{align}
\bar{\Lambda}(U_{B}\ket{\psi}) & =\begin{cases}
\ket{0}\ket{\circ} & \text{if}\;\alpha\alpha^{*}aa^{*}>\beta\beta^{*}bb^{*}\\
\ket{1}\ket{\circ} & \text{if}\;\alpha\alpha^{*}aa^{*}<\beta\beta^{*}bb^{*}
\end{cases}
\end{align}
The outcome of the scattering process depends on both, the amplitudes
of the qubit in the preferred basis and the unknown incoming photon
state.

We do not have any information about the state of the photon, but
we know that it is in a state that is fully described by $\alpha$
and $\beta$. Total ignorance implies that the statistical distribution 
of $\alpha$ and $\beta$ does not depend on the choice of a basis, or in 
other words, the distribution of $(\alpha,\beta)$ must be invariant 
under $SU(2)$ transforms.

One possible distribution%
\footnote{We do not require a normalized photon state and the \emph{linear} process 
is transparent under the choice of the global magnitude. It follows, that the 
distribution of the magnitude does not make a difference as long as the $SU(2)$
symmetry is realized. Hence, we have the freedom of choosing a distribution that
implements the ``directional'' distribution and keeps the calculations simple.%
} that realizes this symmetry is given by
\begin{align}
\alpha & :=G_{1}+iG_{2}\\
\beta & :=G_{3}+iG_{4}
\end{align}
Where $G_{n}$ are mutually independent identically distributed Gaussian
random variables with zero mean. Because the sum of two independent
Gaussian variables results in a new Gaussian variable, this construction
is invariant under unitary transformations. This guarantees $SU(2)$ uniformity
of the distribution. The case $\alpha=\beta=0$ leads to an invalid state, but is of 
zero measure, so we can safely ignore it. The \emph{magnitude} of a complex gaussian 
random variable is Rayleigh distributed, so that we have
\begin{align}
|\alpha| & =R_{1}\\
|\beta| & =R_{2}
\end{align}
with two equally distributed, but independent Rayleigh variables $R_{1}$
and $R_{2}$. The probability density function of each is:
\begin{align}
f(x) & =x\exp(-\frac{x^{2}}{2})
\end{align}
The probability $p\left(|a|\,|\alpha|>|b|\,|\beta|\right)$ can
be expressed in terms of this probability density function:
\begin{align}
p\left(|a|\,|\alpha|>|b|\,|\beta|\right) & =\int_{0}^{\infty}\int_{0}^{\frac{|a|}{|b|}x_{1}}f(x_{1})f(x_{2})dx_{2}dx_{1}\\
 & =\frac{|a|^{2}}{|a|^{2}+|b|^{2}}
\end{align}
And of course, for the complementary event we find:
\begin{align}
p\left(|a|\,|\alpha|<|b|\,|\beta|\right) & =\frac{|b|^{2}}{|a|^{2}+|b|^{2}}
\end{align}
The local observation is therefore
\begin{align}
U_{B}\,:\, a\ket{0}+b\ket{1} & \mapsto\begin{cases}
\ket{0} & \text{with }\, p_{\text{0}}=\frac{|a|^{2}}{|a|^{2}+|b|^{2}}\\
\\
\ket{1} & \text{with }\, p_{1}=\frac{|b|^{2}}{|a|^{2}+|b|^{2}}
\end{cases}
\end{align}
which constitutes the \emph{Born rule}.

This process realizes two key aspects of the measurement postulate.
The outcome locally appears to be a pure state, randomly selected
from a preffered set of orthogonal states. And the probability of 
seeing a specific outcome is proportional to the squared magnitude 
amplitude of the possible outcome states.

We do not yet have the entire measurement postulate however. A complete
derivation must deal with robustness under repeated ``measurements'' and
systems larger than a single qubit. The next section will resolve these
two remaining issues.

\section{Local Quantum Reality}

\subsection{Emergent reality and disrupted time}

In the previous section we have been able to demonstrate the emergence
of the Born rule from the local reconstruction of the behavior of an 
evolving global state. The analysis of the scattering processes has been restricted
to a single qubit in an otherwise unpopulated observer regime and
an initially pure quantum state.

Unfortunately, after a single scattering pass 
the state will not be subjectively pure anymore, but instead consist of two 
branches, encoded by the eigenvectors of the reduced state. A second pass will
again map to the \emph{only two} possible local output branches and the eigenvalues
will mix between the two previous branches. We cannot expect the Born
rule to hold under these conditions. And it turns out, that it does not
in general.

A minor modification to the local system takes care of this problem
and also makes our assumptions more realistic. An observer, who wants
to test for the validity of the Born rule, has to be able to compare sequential
observation results. Consequently, he has to have access to a recorded history
of events. Adding a memory device to the local state of the observer makes
the different outcomes of branches belonging to different iterations
distinguishable and keeps them from mixing. The number of
branches therefore doubles with each \emph{recorded} qubit scattering process.

\begin{figure}
\begin{centering}
\includegraphics[width=1\columnwidth]{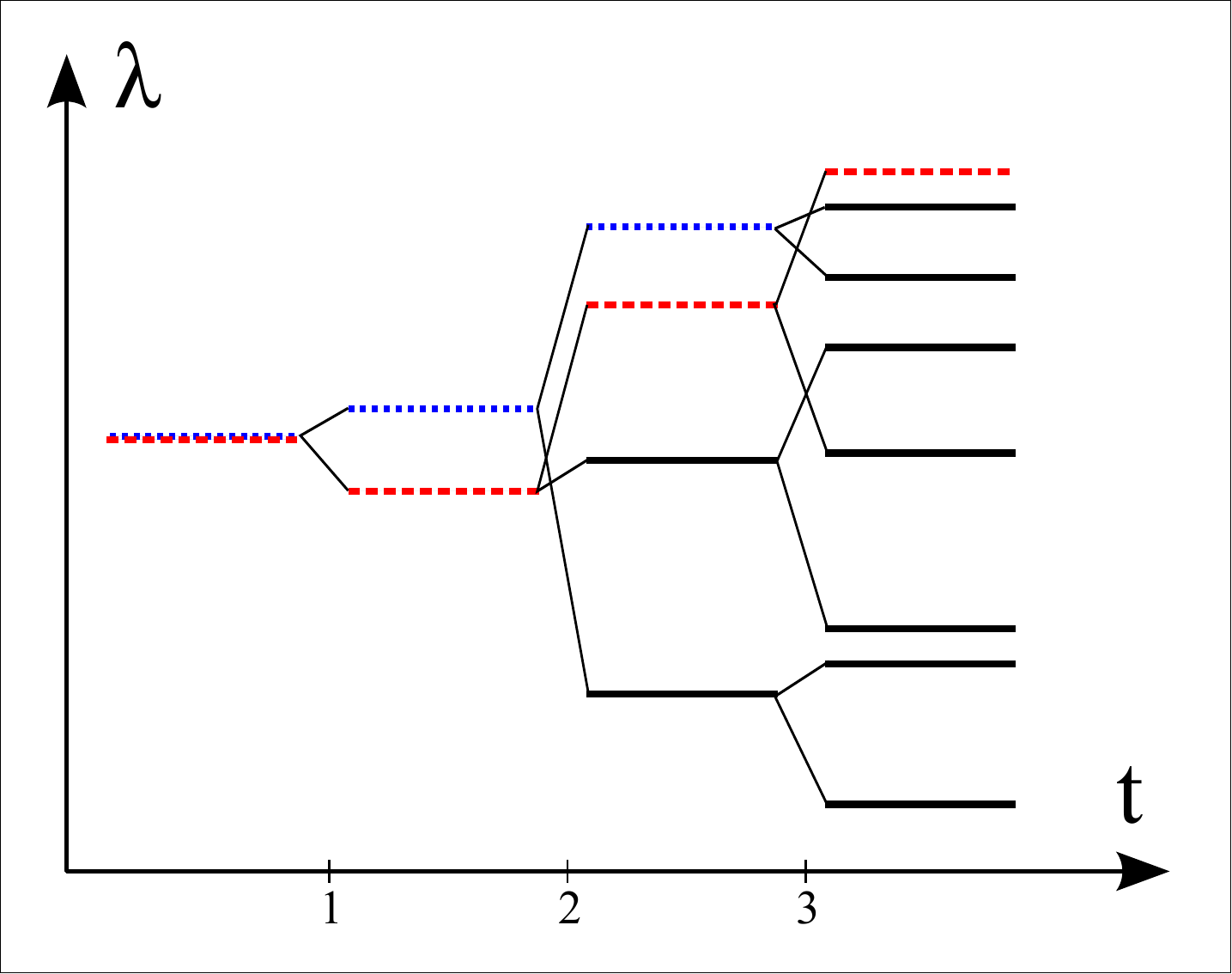} 
\par\end{centering}

\caption{\label{fig:BranchEvolution}Evolution of the branches and their eigenvalues
under Born scattering. The scattering events occur at times 1, 2 and
3. They result in new branches containing the records of the previous
history the respective branch. The dotted blue lines show the dominant
branch and its history after the first and second scattering events.
One event later the red dashed line becomes the dominant branch with
its own history. The single initial branch is part of both histories.}
\end{figure}

Figure \ref{fig:BranchEvolution} shows a sequence of Born scattering
processes and the resulting change of the reconstructable branch, which the
local observer is aware of. Looking only at the sequence of dominant
branches, one can verify that the Born rule holds for the transitions
1 and 2, because the new dominant branch is created from the previous
dominant branch. The situation is identical to the elementary scattering
event with a single branch initial state. That does not hold true
for the third scattering process. The formerly suppressed branch splits
into the new dominant branch. Depending on the actual states, this
sequence of branches usually \emph{breaks} the Born rule.

Despite the partly compromised Born rule for the sequence of reconstructed
states, the observer will \emph{never} see any results, that are in 
disagreement with the Born rule. The reason for this lies in the way the
observer tests the statistics of the results. He keeps a list of old
results to compare to the new ones, and the branch switch not only determines 
the current observed result, but also the list of remembered states. This is also
illustrated in figure \ref{fig:BranchEvolution}. After the third
branch switch, the observer remembers the history marked with the dashed
red line. And this history describes a sequence of observation outcomes,
which is in agreement with the Born rule.

Everything, that the observer would consider as part of his reality
is contained in the current reconstructed state of the universe and
his memory. We can therefore call the sequence of branches, which leads
to the current reconstructed branch and that is stored in the memory
of the observer an emergent subjective reality. A branch switch will
not only change the current perception, but also the perception of
history leading up to this state. The observer can switch between
realities without even noticing, because all records will agree with
the newly formed reality. This picture leaves subjective time and
history disrupted by observed scattering events. This is a drastic,
but seemingly unavoidable consequence of observation. A rough estimation
of probabilities however stongly suggests, that an event, which uncovers a long
time hidden reality branch is extremely unlikely. So the observer's subjective
history is stable after only a few scattering events.

\subsection{Macroscopic interactions and quantum measurement}

The three elementary scattering processes with the locally perceived non-unitary
outcome as described here all act on single qubit systems. While their
mechanisms are very similar, Born scattering comes with the unique
property of scalability to macroscopic systems, as we will discuss now.

Consider a set of projectors
\begin{align}
\mathbb{P} & =\{P_{n}\,:\, n\text{\ensuremath{\in\mathbb{N}}}\}
\end{align}
acting on the Hilbert space $\mathcal{H}_{P}$ with the constraint,
that all their commutators vanish: 
\begin{align}
[P_{m},P_{n}] & =0
\end{align}
We also define the identity operator $\mathbb{I}_{P}$ on $\mathcal{H}_{P}$
and the complementary projectors 
\begin{align}
\bar{P}_{n} & =\mathbb{I}_{P}-P_{n}
\end{align}
We call this set of projectors \emph{complete}, if there is a \emph{special}
orthonormal basis $\{\ket{k_{P}}\}$ of $\mathcal{H}_{P}$ and we
can find two (disjoint) sets $I_{1}(k),I_{2}(k)\subset\mathbb{N}$, so that
\begin{align}
\ket{k_{P}}\bra{k_{P}} & =\prod_{n\in I_{1}(k)}P_{n}\prod_{m\in I_{2}(k)}\bar{P}_{m}
\end{align}
for all $k$. The set is \emph{independent}, if we cannot remove any
projectors from $\mathbb{P}$ without giving up completeness.

With the set of projectors, there is a natural way to define a unitary
evolution to split up vectors in $\mathcal{H}_{P}$.
\begin{align}
U_{n} & : & \ket{\psi} & \longmapsto P_{n}\ket{\psi}\otimes\ket{0}+\bar{P}_{n}\ket{\psi}\otimes\ket{1}
\end{align}

Here, $\{\ket{0},\ket{1}\}$ is the basis of a qubit%
\footnote{When these evolutions are concatenated, the qubit must be assumed
to be a different one in each stage. This is not reflected in our
notation, in order to keep it simple.%
} and we can apply the Born scattering evolution to it, with the local
result
\begin{align}
\bar{\Lambda}(U_{B}U_{n}\ket{\psi})= & \begin{cases}
P_{n}\ket{\psi}\ket{0} & \text{with }p=\frac{\braket{\psi|P_{n}|\psi}}{\braket{\psi|\psi}}\\
\\
\bar{P}_{n}\ket{\psi}\ket{1} & \text{with }p=\frac{\braket{\psi|\bar{P}_{n}|\psi}}{\braket{\psi|\psi}}
\end{cases}
\end{align}
We have seen, that the recorded subjective observation of qubits is
consistent and stable, allowing us to restrict our discussion to the
dominant branch and taking the position of the local observer. Repeating
the observation with a different projector $P_{m}$ and a \emph{fresh qubit}
maps the first branch to $P_{m}P_{n}\ket{\psi}\ket{0}\ket{0}$ with
the old qubit state as the last factor. For a cleaner notation, we
write the qubit state ordered list inside a single ket $\ket{0,0}$.
The probability of finding this branch combines the probabilities
from both scattering events and results in:
\begin{align}
p & =\frac{\braket{\psi|P_{n}^{\dagger}P_{m}^{\dagger}P_{m}P_{n}|\psi}}{\braket{\psi|P_{n}|\psi}}\cdot\frac{\braket{\psi|P_{n}|\psi}}{\braket{\psi|\psi}}\label{eq:BornRuleCancel}\\
 & =\frac{\braket{\psi|P_{m}P_{n}|\psi}}{\braket{\psi|\psi}}
\end{align}

As can be seen from generalizing this calculation, further scatterings
only result in adding more projectors to the numerator of the probability
expression. The order is arbitrary, because the projectors commute
by definition. This is also true for the projectors in front of the
state $\ket{\psi}$. We can therefore choose the canonical ordering
of the index without changing the properties of the result, as far
as they relate to the state $\ket{\psi}$. The order of the qubit
history will change however. This motivates the definition of the
operator
\begin{align}
M & =\prod_{n=1}^{N}U_{B}U_{n}
\end{align}
where each factor comes with a fresh qubit. We also define $\ket{[j]}$
as the qubit list with the $N$-digit binary expansion of $j$. Similarly,
we define $P_{[j]}$ to be the product sequence with the digits $\{P_{n},\bar{P}_{n}\}$
 following the $N$-digit binary representation of $j$. 

The subjective local result of the application of $M$ on the state
$\ket{\psi}$ is then:

{\footnotesize 
\begin{align}
\bar{\Lambda}(M\ket{\psi}) & =\begin{cases}
P_{[0]}\ket{\psi}\ket{[0]} & \text{with }p=\frac{\braket{\psi|P_{[0]}|\psi}}{\braket{\psi|\psi}}\\
P_{[1]}\ket{\psi}\ket{[1]} & \text{with }p=\frac{\braket{\psi|P_{[1]}|\psi}}{\braket{\psi|\psi}}\\
\vdots & \vdots\\
P_{[2^{N}-2]}\ket{\psi}\ket{[2^{N}-2]} & \text{with }p=\frac{\braket{\psi|P_{[2^{N}-2]}|\psi}}{\braket{\psi|\psi}}\\
P_{[2^{N}-1]}\ket{\psi}\ket{[2^{N}-1]} & \text{with }p=\frac{\braket{\psi|P_{[2^{N}-1]}|\psi}}{\braket{\psi|\psi}}
\end{cases}
\end{align}
}{\footnotesize \par}

We are interested in the limit of very large $N$. In this case, the $P_{[k]}$
contain all possible projector lists as sublists. For a given list
of projections, that multiplies to a projector on a single dimensional
subspace, we can find a \emph{unique} binary sequence, which produces
this list as a sublist and preserves that subspace in the remaining
projectors, because either $P$ or $\bar{P}$ preserves the subspace.
All other lists containing the same sublist must multiply to $0$.

The consequence is, if we have a \emph{complete} set of projectors
$\mathbb{P}$, then there is exactly one outcome with non-zero probability
resulting in the state $\ket{k_{P}}\otimes\ket{[j]}$, with a probability
of 
\begin{align}
p & =\frac{\braket{\psi|k_{P}}\braket{k_{P}|\psi}}{\braket{\psi|\psi}}
\end{align}
Summarized, for \emph{complete} $\mathbb{P}$ and sufficiently large
$N$, the scattering iteration
\begin{align}
\bar{\Lambda}(M\ket{\psi}) & =\ket{k_{P}}\otimes\ket{[j]} & \text{with } & p=\frac{\left|\braket{k_{P}|\psi}\right|^{2}}{\braket{\psi|\psi}}
\end{align}
results in the \emph{Born rule} for measurement in the basis $\{\ket{k_{P}}\}.$

We have constructed a measurement mechanism, that works for measured 
Hilbert spaces of arbitrary size and can therefore be applied to 
macroscopic systems. The mechanism does not need very special initial
condition or careful tuning. In fact, it is very robust, as the order 
of the elementary scattering processes does not change the outcome, nor 
does the actual choice of projectors. A physical device realizing this
mechanism should not be hard to design and build, given the single qubit
scattering process can be realized.

So far, we only discussed the case of a \emph{complete} set of projectors.
The framework presented allows for generalizations, that are not
discussed here however. It is interesting to note, that the Born rule
is the only rule, which delivers robust and consistent results for a
macroscopic system built from single qubit interactions, mostly due
to the canceling terms in equation \eqref{eq:BornRuleCancel}. 

\subsection{Realizing Born scattering on a qubit}

The Born scattering process presented in the last section is very generic
and does not refer to a specific physical system. As an example of
a simple system, that can realize the scattering process, we will discuss
the interaction between a single electron bound inside an atom%
\footnote{This model can be extended to several bound electrons, 
but we want to keep it simple here}
and its interaction with the electromagnetic radiation field.
We are looking at a transition of the electron, that happens between
two energy levels and also changes the angular momentum. For example, 
in terms of the standard hydrogen quantum numbers, the states could be $\ket{n=1,l=0}$ 
and $\ket{n=2,l=1}$. For a simpler notation, we use the qubit notation
$\ket{0}$ and $\ket{1}$ respectively.

With the angular momentum eigenstates for the qubit, the incoming single
photon is best described in terms of a circular polarization basis
$\ket{\rightsquigarrow\circlearrowleft}$ and 
$\ket{\rightsquigarrow\circlearrowright}$. 
The arrow to the right merely indicates the direction of the photon 
trajectory towards the atom prior to the collision.

If the incoming photon carries the wrong angular momentum for the 
transition, then nothing happens:

\begin{align}
\ket{\rightsquigarrow\circlearrowleft}\ket{0} \mapsto \ket{0}\ket{\rightsquigarrow\circlearrowleft}
\end{align}

The photon with the opposite angular momentum can either be absorbed while exciting 
the electron, or shake the system a little and radiate away in a different 
direction.\footnote{Or more precisely, in a spatial scattering state.}
We assign the same amplitude\footnote{ This is a necessary assumption
at this point. I do however believe, that it is possible to derive it from
a proper interaction model.} to both options. The relative phase 
between them is irrelevant, it cancels in the reduction process.

\begin{align}
\ket{\rightsquigarrow\circlearrowright}\ket{0} \mapsto \left( \ket{1}\ket{\circ} + \ket{0}\ket{\leftrightsquigarrow\circlearrowright}\right)\frac{1}{\sqrt{2}}
\end{align}

Here, $\ket{\circ}$ is the relative 0-photon state and $\ket{\leftrightsquigarrow}$
is the state of the scattered photon, which is orthogonal to the unchanged outgoing
photon state.

Trying to add another angular momentum quantum to the already excited
state also has to leave the system unchanged:

\begin{align}
\ket{\rightsquigarrow\circlearrowright}\ket{1} \mapsto \ket{1}\ket{\rightsquigarrow\circlearrowright}
\end{align}

However, if we take the angular momentum of the excited state back by sending in
and oppositely polarized photon, we can again have two possible effects. The 
incoming photon can either trigger the emission of a second photon with the same properties.
Or the incoming photon only shakes the atom slightly and leaves in a new direciton.
As above, the amplitudes of both processes are assumed to be equal while relative
phase is not of importance.

\begin{align}
\ket{\rightsquigarrow\circlearrowleft}\ket{1} \mapsto \left( \ket{0}\ket{\rightsquigarrow\circlearrowleft,\rightsquigarrow\circlearrowleft} + \ket{1}\ket{\leftrightsquigarrow\circlearrowleft} \right)\frac{1}{\sqrt{2}}
\end{align}

These four maps between an orthonormal basis of the input space and
an orthogonal basis of a significantly different output space together
describe a unitary evolution, that matches all the criteria, we have 
identified earlier for generating the Born rule for a local observer.

If it can be shown, that the equal amplitude assumption for the scattering 
outcome states is in agreement with light-matter interaction theory, then
we should expect to find this mechanism everywhere around us.

\section{Conclusion}

We have presented a possible solution to the quantum measurement problem, that
does not require any modification of the dynamics or state space of traditional
quantum theory. Nor does it demand conditions for classicality or a specific
kind of observer. The results are derived solely from a realist understanding 
of the state of the universe and the assumption, that identically behaving systems
are in fact physically identical.

One of the main results is, that any internal description of the quantum universe leads
to a \emph{non-linear} model of the state of the universe and its evolution.
This description can change \emph{discontinuously} upon the interaction with ambient photons.
We have seen, that, under certain conditions, the projective and random outcome, as
described by the measurement postulate, emerges naturally. The corresponding
mechanism and its consequences are within reach of experimental verification.

The derived measurement process nicely generalizes to arbitrarily large systems. A POVM
\citep{Audretsch2007}
description of measurements is also covered, in the form of projective measurements on larger
spaces. Furthermore, we have given a generalization of the concept of a quantum subsystem
 \citep{Neumann1927}, that does no longer have to be a tensor-factor space of the 
state space.

The derived results are compatible with relativistic quantum theory.\footnote{While 
we do not provide a full QFT formulation, the general principles translate directly
from the relativistic Fock space to the domain of quantum field theory.} In fact, they require
a relativistic universe to be deducible. At the same time, the physicality of
the nonlocality of the quantum state space is revealed, that is normally hidden under the inderterminism of observations.
The underlying breach of Lorentzian symmetry might become experimentally accessible, if deterministic
control of measurement outcomes\footnote{The measurement mechanism suggests, that controlling
the polarization of a collapse-inducing photon makes this possible.} can be physically realized.

The key idea of physical behaviorism does not play a relevant role in any
main stream interpretation\footnote{What we present here is a theory with explanatory
power and predictions. But it competes with \emph{interpretations}, mostly on issues of
ontology.} of quantum theory. While this fact alone makes it stand out, there are still
many more fundamental differences. A great advantage is, that the locally perceived pure state
of the universe is very robust\footnote{There is no preferred basis problem. The
eigenstructure of localized states is independent of the choice of a description.}
and well defined, unlike in theories, that exclusively rely on entanglement to
define features of histories, events and relationships. Examples for these
interpretations include MWI \citep{Everett1957,DeWitt1973}, consistent histories 
\citep{Dowker1996,Griffiths1984,Okon2013}, but also bare decoherence 
\citep{Zeh1993,Zurek2002,Schlosshauer2007,Joos2003} and decoherent histories \citep{Bassi1998}.

Everett's relative state interpretation takes the idea of subjective reality very far.
Not only does every observer experience his own reality, each observer also splits up
into a vast number of copies, who also experience all possible histories \citep{Everett1957}.
The theory we present here leads to a much more restricted notion of subjective reality.
A single observer's perception of reality is greatly restricted by what
he does not know about the universe. That is foremostly the state of incoming or
outgoing photons. Two observers at roughly the same position in spacetime will share
practically all of their ignorance about these states and therefore come to the same
conclusions about their local reality. By extension and covering a whole time-slice with 
observers, that have spatially overlapping descriptions of their immediate past, we can deduce
the existence of a single connected subjective reality. There is \emph{only one} ``world''. 
And it specifically does \emph{not} depend on the strategy of the observer to deal with it.
Still, we do share the branching structure with MWI and we have seen, that, on a short time
scale, many different realities can be realized, but one at a time. Only a single one of those realities
does have a realized future in the long run, but all do have a full recorded history reaching
into the past.

Ontologically, the given theory of quantum measurement is very attractive. It combines
the objective realism of the universal quantum state with subjective features of
one's own description of this state. All postulates are harmless, from a philosophical
point of view, and they represent well established physics. The notion of an all-creating
dynamical law reduces the ontological requirements of the theory to the absolute
minimum. It also allows to avoid fuzzy concepts, like improper mixtures for the description
of what we know about quantum states.

One mostly philosophical problem does emerge however. The concept of linear time
as a reference for events in reality cannot be maintained from the global perspective
of the subjective reality. Undetectable to the observer, different alternate realities
can fight for becoming the dominant one, at least over a short period of time.
This effect appears to be highly unsettling and not really greatly preferable to the
world-splitting in the Everett  interpretation. Currently, and with the evidence
presented here, there does not seem to be a way around it. Further research will have
to analyze this in more depth and may offer alternatives.

The local observation theory uniquely\footnote{Theories like GRW \citep{Ghirandi1986},
which use modifications of the linear evolution, may share that feature.
The uniqueness refers to interpretations and theories, that do not require
fundamental changes in the structure of QT.} features a physical mechanism 
with an objective source\footnote{The photon polarization.} of randomness. 
That means unlike MWI, no additional probability structure is needed for us to 
be able to consistently speak about the frequency\footnote{Notably, we also do not have
to refer to Bayesian probabilities.} of observed events. Also unlike Bohmian 
mechanics \citep{Bell1982,Bell2004}, there are no fragile initial conditions
to be met in order to recover the Born statistic.\footnote{Although it should be mentioned,
that the scattered photons have to be at least locally separable from the target. That is
a constraint on the initial conditions on the universe, however not a very demanding one.
Much more, it is a definition of an arrow of time.}

A certain detail in the derivation Born rule needs some further attention. The assumption
of equal amplitude magnitudes for the two concurring scattering processes is not obviously
true. The symmetry of the choice and the existence of certain conservation laws seem to
favor the equality, but it is by no means guaranteed to hold. This is an area, which will
need more investigation and a better model. The predictive power of this choice
is nevertheless so great, that the assumption can be just justified for now. This is
particularly true with the Born rule being the only statistical process, that survives
macroscopification, bringing in the aspect of natural selection for observable 
processes.

The proposed solution significantly differs in all relevant aspects from the established
interpretations of quantum theory and is at the same time experimentally accessible. Even
more, it has the potential to make the nonlocal nature of quantum theory directly available
for experiments. The possibility of controlling subjectively indeterministic
processes by manipulation of the underlying scattering processes could greatly
extend the tool set of quantum manipulations and forces us to rethink established results%
\footnote{Like no-cloning \citep{Zurek1982} or no-signalling}, that rely on the 
fundamentality of randomness in quantum observations or linearity.

\section*{Acknowledgments}
\begin{acknowledgments}
I would like to thank David Pringle for the ongoing financial and motivational support
Many people contributed to my thoughts by discussing quantum theory with me on different
levels. Thank you all for being so inspiring. I am also specifically very thankful to
Michael Nock for the enriching discussions about the ideas presented here.
Finally, I would like to express my gratitude to Juergen Audretsch for having been an excellent
teacher, who helped me to stay curious and open minded.
\end{acknowledgments}

\bibliographystyle{plain}
\bibliography{QuantumReferences}

\end{document}